\documentclass[a4paper,11pt]{article}
\usepackage{url} 
\usepackage{graphicx,subcaption}
\usepackage{jinstpub} 

\usepackage{hyperref}
\hypersetup{colorlinks,breaklinks,
            linkcolor=blue,urlcolor=blue,
            anchorcolor=blue,citecolor=blue}

\newcommand{\onbb}      {$0\nu\beta\beta$}

\newcommand{\LEG}       {LEGEND}
\newcommand{\Ltwo}      {{\LEG-200}}

\newcommand{\be}        {\begin{equation}}
\newcommand{\ee}        {\end{equation}}

\usepackage{siunitx}
\usepackage{romannum}
\usepackage{xcolor}

\title{Calibration sources for the LEGEND-200 experiment}

\author[a]{L.~Baudis}
\author[b]{G.~Benato}
\author[c]{E.~M.~Bond}
\author[a]{P.-J.~Chiu}
\author[c]{S.~R.~Elliott}
\author[c]{R.~Massarczyk}
\author[c]{S.~J.~Meijer}
\author[a]{Y.~M\"uller}

\affiliation[a]{University of Zurich, 8057 Zurich, Switzerland}
\affiliation[b]{Gran Sasso Science Institute, 67100 L'Aquila, Italy}
\affiliation[c]{Los Alamos National Laboratory, 87545 Los Alamos, New Mexico, USA }
\emailAdd{massarczyk@lanl.gov, yannick.mueller@physik.uzh.ch, pin-jung.chiu@physik.uzh.ch}

\abstract{In the search for a monochromatic peak as the signature of neutrinoless double beta decay an excellent energy resolution and an ultra-low background around the $Q$-value of the decay are essential. The LEGEND-200 experiment performs such a search with high-purity germanium detectors enriched in $^{76}$Ge immersed in liquid argon. To determine and monitor the stability of the energy scale and resolution of the germanium diodes, custom-made, low-neutron emission $^{228}$Th sources are regularly deployed in the vicinity of the crystals. Here we describe the production process of the 17 sources available for installation in the experiment, the measurements of their alpha- and gamma- activities,  as well as the determination of the neutron emission rates with a low-background LiI(Eu) detector operated deep underground. With a flux of $\left( 4.27 \pm 0.60_{\rm stat} \pm 0.92_{\rm syst} \right) \times 10^{-4} ~\text{n / (kBq$\cdot$s)}$, approximately one order of magnitude below that of commercial sources, the neutron-induced background rate, mainly from the activation of $^{76}$Ge, is negligible compared to other background sources in LEGEND-200.}

\begin{document}
\maketitle
\flushbottom

\section{Introduction}
\label{sec:intro}

Two outstanding questions in particle physics are the absolute mass scale of neutrinos, and their nature, namely whether they are Majorana or Dirac particles. The observation of $0\nu\beta\beta$ decay would demonstrate that neutrinos are Majorana fermions and would indicate violation of total lepton number conservation.
For light Majorana neutrino exchange, the predicted rate is proportional to the effective Majorana neutrino mass, $m_{\beta\beta} = |\Sigma_i U^{2}_{ei}m_{\nu_i}|$, where the sum is over the mass eigenstates $m_{\nu_i}$, and $U_{ei}$, the corresponding entries in  the lepton mixing matrix, are complex numbers. 
The experimental observable in $0\nu\beta\beta$ decay is the half-life, $(T_{1/2})^{-1}\propto m^2_{\beta\beta}$, which current estimates and experimental constraints indicate to be in the range between T$_{1/2}\sim10^{26}-10^{28}$\,years for any neutrino mass ordering and beyond $10^{28}$\,years for the normal ordering only~\cite{Workman:2022ynf}. 
Among the various techniques and isotopes employed to search for this rare decay, those using high-purity germanium detectors enriched in $^{76}$Ge, such as GERDA~\cite{Gerda:2020xhi} and the \textsc{Majorana Demonstrator}~\cite{Majorana:2019nbd}, have reached unprecedented half-life sensitivities, which is notable due to their relatively small exposure compared to the alternative approaches. This is due to their ultra-low backgrounds around the $Q$-value of 2039\,keV, their excellent energy resolutions, high isotopic abundance, and high efficiency for the detection of the two emitted electrons. The LEGEND experiment~\cite{LEGEND:2021bnm}, building upon the successes of its predecessors, aims to construct a ton-scale experiment to search for the  $0\nu\beta\beta$ decay of $^{76}$Ge with a half-life sensitivity beyond 10$^{28}$\,years, corresponding to an upper limit on $m_{\beta\beta}$ in the range of 9–21\,meV, depending on nuclear matrix element, in 10\,ton-years of exposure.

The first, 200-kg phase of LEGEND, called LEGEND-200, is in commissioning phase at the Laboratori Nazionali del Gran Sasso (LNGS) in Italy. LEGEND-200 will deploy the existing 70\,kg of enriched detectors from GERDA and the \textsc{Majorana Demonstrator}, together with an additional 130\,kg of newly produced p-type inverted-coaxial point-contact detectors~\cite{Cooper:2011abc} in liquid argon, in an upgraded GERDA infrastructure. With anticipated initial science runs beginning in late 2022, the goal of LEGEND-200 is to reach a half-life sensitivity of 10$^{27}$\,years after five years of operations. 

To achieve its science goal, an excellent energy resolution and stability of the energy scale is fundamental. To this end, $^{228}$Th calibration sources are regularly deployed in the liquid argon, close to the germanium detector arrays. One crucial requirement for the radioactive sources is a low neutron emission rate, to avoid activation of the Ge crystals and surrounding materials, in particular the production of the $^{77}$Ge isotope~\cite{Wiesinger:2018qxt,Pandola:2007hv}. In this manuscript, we describe the requirements, production, and characterization of the calibration sources for LEGEND-200 in detail. 

\section{Low-background calibration sources}
\label{sec:low_bakground_calibration_sources}

To regularly monitor the stability of the signal gain, energy scale, and energy resolution of the germanium detectors, as well as the efficiency of the liquid argon veto system, calibration using radioactive sources is necessary. The $^{228}$Th isotope was selected as the source for LEGEND-200 because of the following reasons. 
$^{228}$Th undergoes multiple $\alpha$ and $\beta$ decays before reaching the stable $^{208}$Pb isotope. Decays to excited states of daughter nuclei subsequently generate monoenergetic $\gamma$ rays, where a substantial number of high-statistics $\gamma$ lines between \num{500} and $\SI{2600}{keV}$ are available for calibration up to the $Q$-value of the $0\nu\beta\beta$ decay, $Q_{\rm \beta\beta} = \SI{2039}{keV}$. 
In addition, the topology of the $\gamma$ lines are useful for the pulse shape analysis for germanium detectors. The dominant $\gamma$ line with the highest energy in the decay chain of $^{228}$Th to $^{208}$Pb is the full energy peak of $^{208}$Tl at \SI{2614.5}{keV}. This high-energetic $\gamma$ ray may undergo $e^{+}e^{-}$ pair production in the germanium detectors, where the electron deposits energy in the detector while the positron annihilates with an atomic electron in the crystal and creates two annihilation photons. A double-escape event is a scenario when both annihilation photons escape from the detector, leaving a signature that is topologically equivalent to the expected $0\nu\beta\beta$ signal; therefore, double-escape peaks (DEP) are used to calibrate the pulse-shape-discrimination (PSD) technique. The DEP of $^{208}$Tl at $(2614.5-2\times511)\,\mathrm{keV} = \SI{1592.5}{keV}$ provides sufficient statistics, making $^{228}$Th an optimal source candidate. 
The more abundant and cheaper $^{232}$Th would also provide the same $\gamma$ lines. However, its progeny $^{228}$Ac emits a $\gamma$ with \SI{1588.2}{keV} of energy, which would partially overlap with the $^{208}$Tl DEP and could spoil the PSD calibration.
$^{228}$Th sources had been used in other $0\nu\beta\beta$ experiments, such as GERDA Phase \Romannum{1}, Phase \Romannum{2} \cite{Baudis2015, Agostini2021}, and the \textsc{Majorana Demonstrator} \cite{Abgrall2017}.   

\begin{figure}[htbp]
\centering
\hspace{1cm}
\includegraphics[width=.225\textwidth]{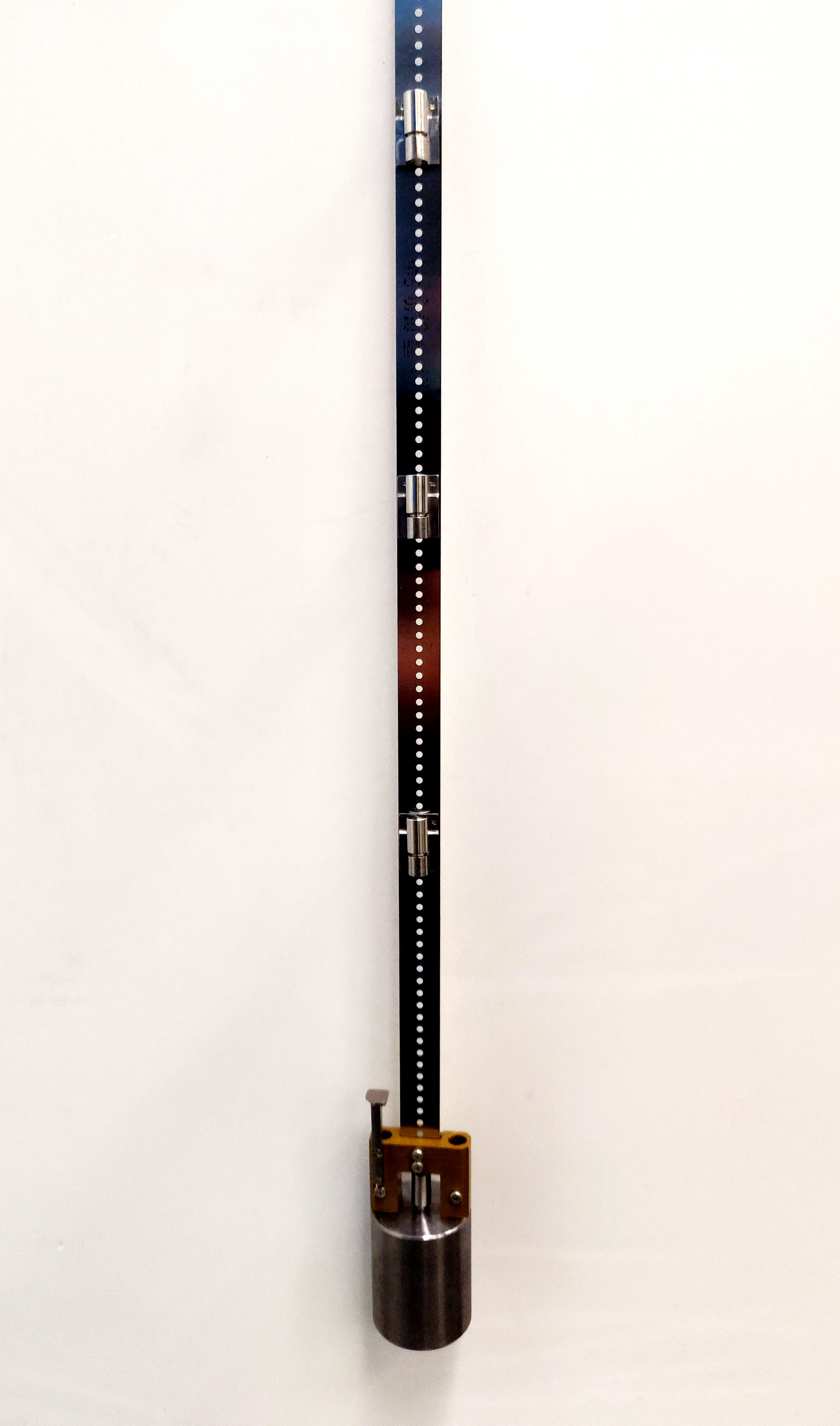}
\qquad
\includegraphics[width=.33\textwidth]{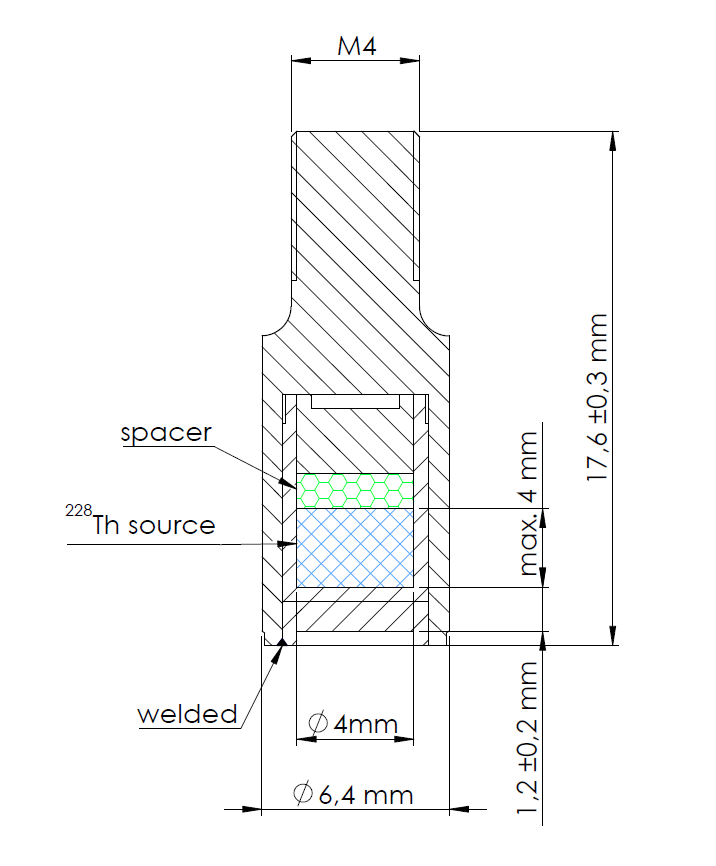}
\caption{\label{Fig_Welding} (Left) Picture of three mock source containers mounted into source holders spot-welded onto a steel band. The source at the bottom is mounted into a thread in a tantalum absorber, which is attached to the band with a Torlon holder. During calibrations, the steel band can be moved via remote control of the motor to which it is connected. (Right) Drawing of the calibration source container. The design follows the GERDA source design.}
\label{fig:SIS}
\end{figure}

To calibrate the germanium detectors, four calibration systems, following the design used in the GERDA experiment \cite{Froborg2012, Baudis2013}, were constructed and deployed. One system consists of a steel band with four $^{228}$Th sources mounted on top (see Fig.\,\ref{fig:SIS} left). The systems are located above the cryostat and the sources are located around \SI{8}{m} above the detector array when not in use during $0\nu\beta\beta$ decay runs. The calibration is foreseen to be performed on a weekly basis with a duration of a few hours. During the calibration, the steel bands are rolled vertically down, bringing the sources into the cryostat to the vicinity of the germanium detectors.
The activity of each $^{228}$Th source is $\sim \SI{5}{kBq}$, which  provides sufficient statistics and uniform coverage for the calibration of the detectors, while not exceeding the limits posed by the signal readout and the data acquisition system \cite{Ransom2021}. This was determined based on simulations with the MaGe framework \cite{Boswell2011, Bauer2006} within the LEGEND collaboration,  and is consistent with the source activities of $\sim \SI{20}{kBq}$ in GERDA Phase \Romannum{2}, where each system comprised one source. The deployment of multiple sources spread along one system ensures simultaneous calibration of multiple detectors with good homogeneity of statistics within a few hours.

A $^{228}$Th nucleus undergoes five $\alpha$ decays before reaching the stable $^{208}$Pb nucleus, emitting $\alpha$ particles with energies in the range of $\num{5.2}-\SI{8.8}{MeV}$. These $\alpha$ particles yield neutron fluxes through $\left( \alpha, n \right)$ reaction, which can lead to background events. On the one hand, these parasitic neutrons in LEGEND-200 can activate $^{76}$Ge, producing $^{77}$Ge (half-life \SI{11.3}{h}) and $^{77 \rm m}$Ge (half-life \SI{53.7}{s}) isotopes, which undergo $\beta$ decays with $Q$-values larger than \SI{2}{MeV}. On the other hand, neutrons can be captured by surrounding materials, producing high-energetic $\gamma$ rays. A solution to reduce the number of neutrons is to embed the $^{228}$Th in materials with energy thresholds of the $\left( \alpha, n \right)$ reaction larger than $\SI{8.8}{MeV}$, the highest energy of the $\alpha$ particles that can be generated in the decay chain of $^{228}$Th to $^{208}$Pb. Gold, which has an $\left( \alpha, n \right)$ energy threshold of $\SI{9.94}{MeV}$, a good ductility, and an ease of procurement, turns out to be the best candidate, as described in \cite{Maneschig2012}. Therefore, gold was selected as the substrate for the source material, to minimize the neutron production.   

\section{Production of radioactive sources}
\label{sec:Production}
The \Ltwo{} design requires the production of sixteen source containers. This section introduces how the source materials are deposited on foils to prevent $(\alpha,n)$ reaction and placed inside the containment shown in Fig.~\ref{fig:SIS}, the containers are sealed by welding, and the seal is tested under cryogenic conditions.

\subsection{Electrodeposition of $^{228}$Th}

\begin{figure}[htbp]
\centering 
\includegraphics[width=.43\textwidth]{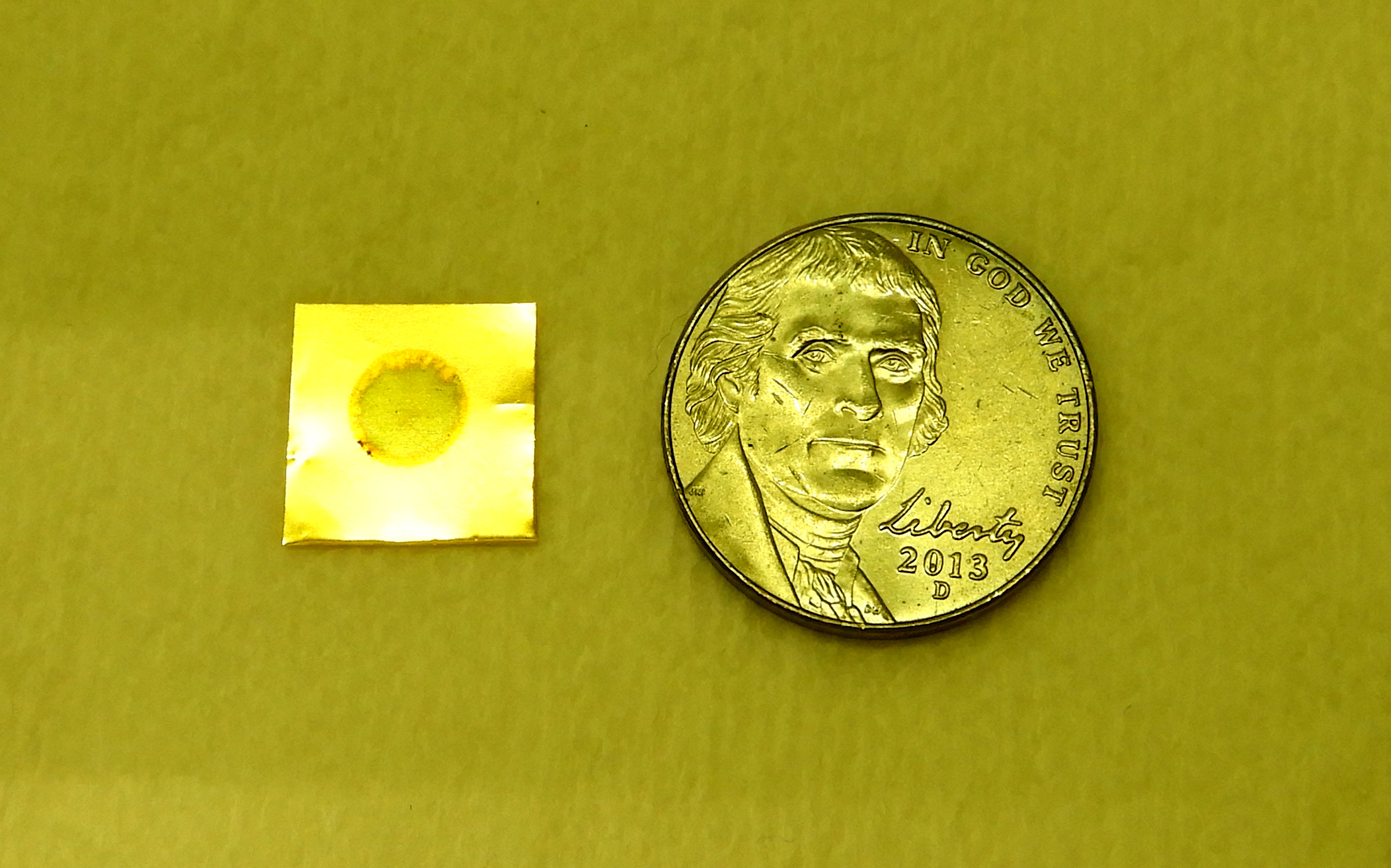}
\qquad
\includegraphics[width=.38\textwidth]{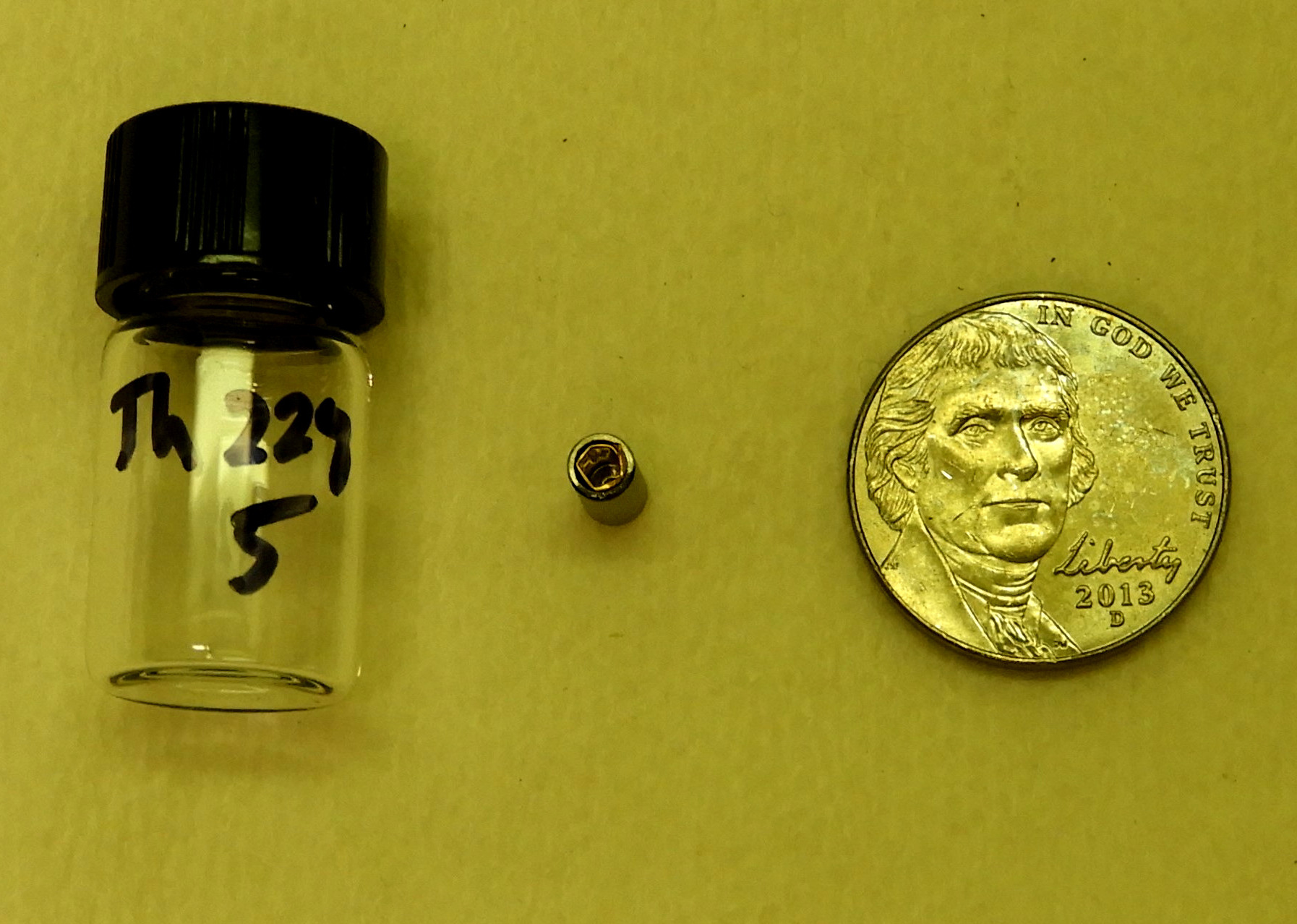}
\caption{\label{Fig_EDeposition} (Left) Deposited thorium on gold foil with US-nickel for size comparison. (Right) The same foil after it was rolled and placed inside the inner cylinder of the source assembly.}
\end{figure}

Radioactive material was acquired from Eckert\,$\&$\,Ziegler\footnote{\url{www.ezag.com}}. 
A total of 97.8\,kBq $^{228}$Th in form of thorium-chloride dissolved in HCl was used to produce the initial set sources to be installed in the LEGEND-200 experiment. 
Via electrodeposition, a thin thorium film was placed on a 50-\si{\micro m}-thin gold foils.
The electrodeposition method used was based on the ammonium sulfate method previously used for uranium electrodeposits \cite{Bond2008}. The $^{228}$Th solution was dried on a hotplate in a small glass, dissolved in pH2 0.5M ammonium sulfate solution, and transferred to the 0.25” glass electrodeposition cell. The $^{228}$Th was deposited from $\sim$1.5\,mL of the plating solution for 30 minutes at 43\,mA.  The gold foils had a purity of 99.995\% and were 1" $\times$ 1" in size. Each deposit was arranged on a separate foil in a circle of about $1/4$" diameter in the center. A first set of depositions was followed by immediate measurements of the $\alpha$ activity of the $^{228}$Th decay using a silicon detector. Due to the fast in-growth of daughter isotopes, it is difficult to estimate the yield. For test samples, between 50 and 82\% yield from solution to deposit was observed. Based on this observation, the amount of $^{228}$Th solution was increased so that the desired activity of around 5\textendash6\,kBq per source was deposited. After the deposition, the external part of the foil with no $^{228}$Th activity was cut off, and the foil rolled. Figure~\ref{Fig_EDeposition} shows the deposited material as well as the folded foil inside the inner-most containment. 

\subsection{Welding and cryogenic tests}

After electrodeposition and placement in the inner containment, the capsules were sealed by electron beam welding under vacuum. Due to the size of the sample, the welding joints were analyzed for inhomogeneities by Metallography, see Fig.~\ref{Fig_Welding}. A small cap was welded to the inner assembly, the assembly placed upside down in the outer containment, and then sealed by welding an end cap. 

\begin{figure}[htbp]
\centering 
\includegraphics[width=.43\textwidth]{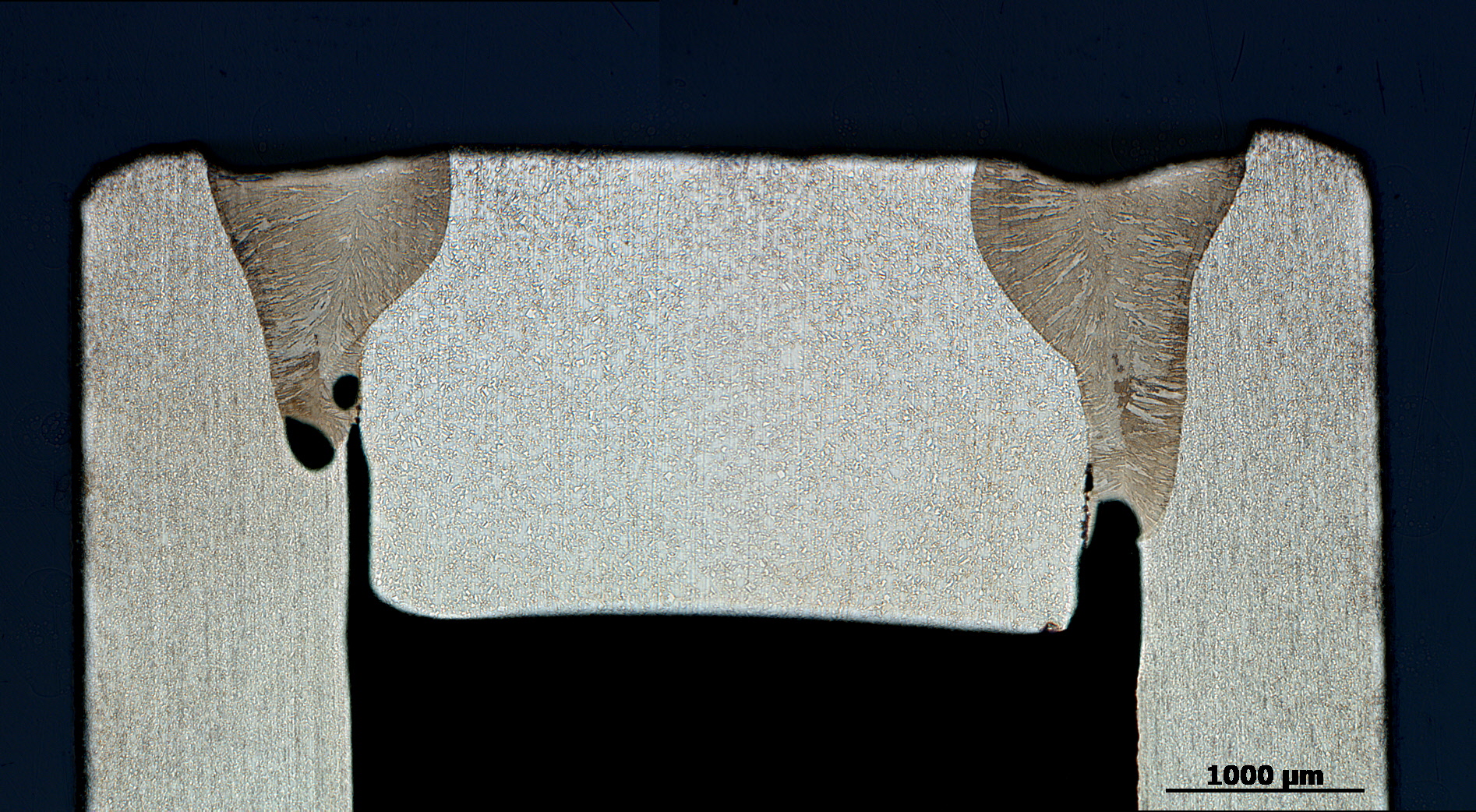}
\qquad
\includegraphics[width=.40\textwidth]{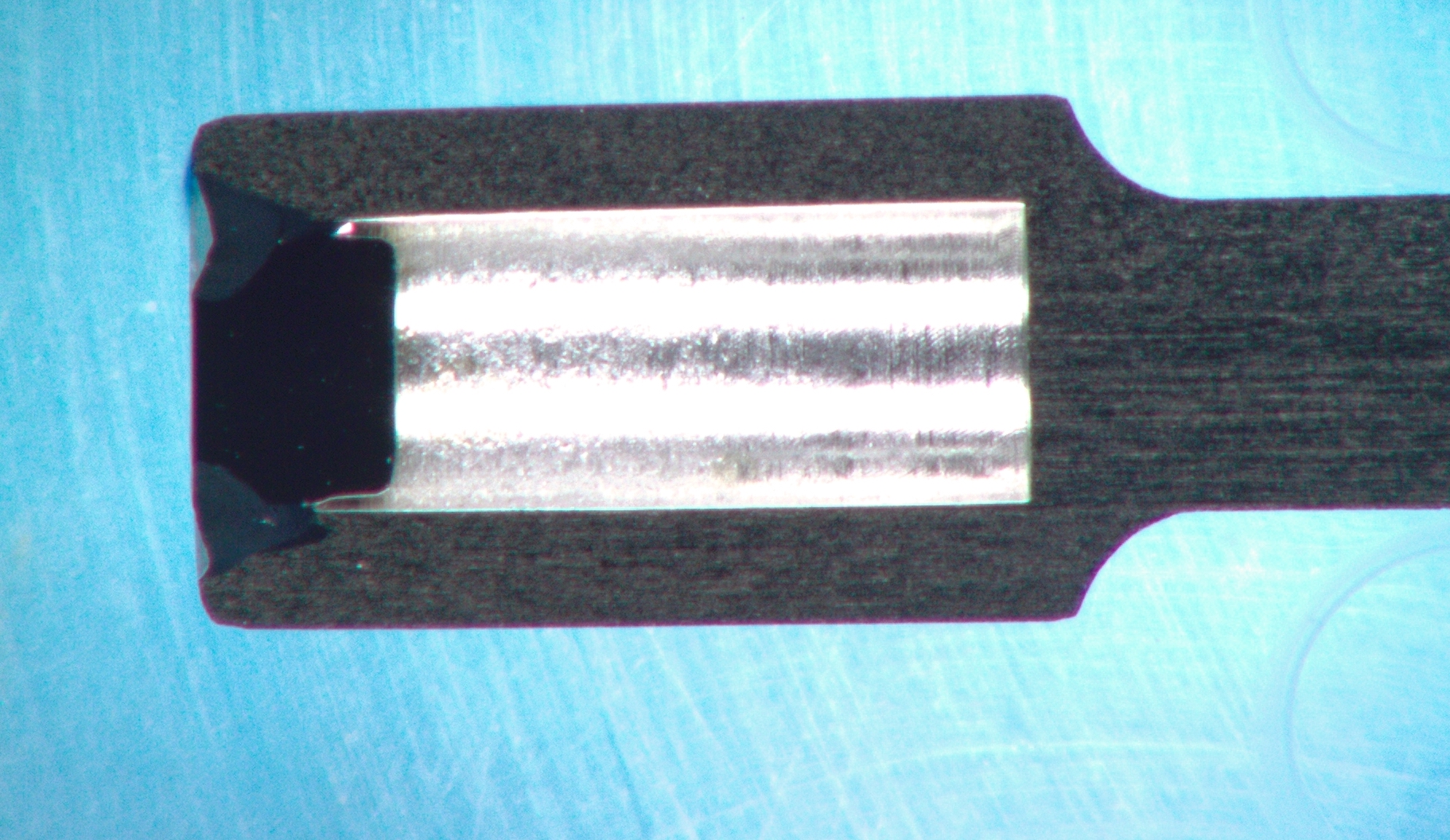}
\caption{\label{Fig_Welding} (Left) Metallography of a welding joint between outer containment and end cap. Different welding settings were tested, so that problems like bubbles (black spots in left joint) are avoided. (Right) Full picture of a cut-open outer source containment after welding tests.}
\end{figure}

\begin{figure}[htbp]
\centering 
\includegraphics[width=.30\textwidth]{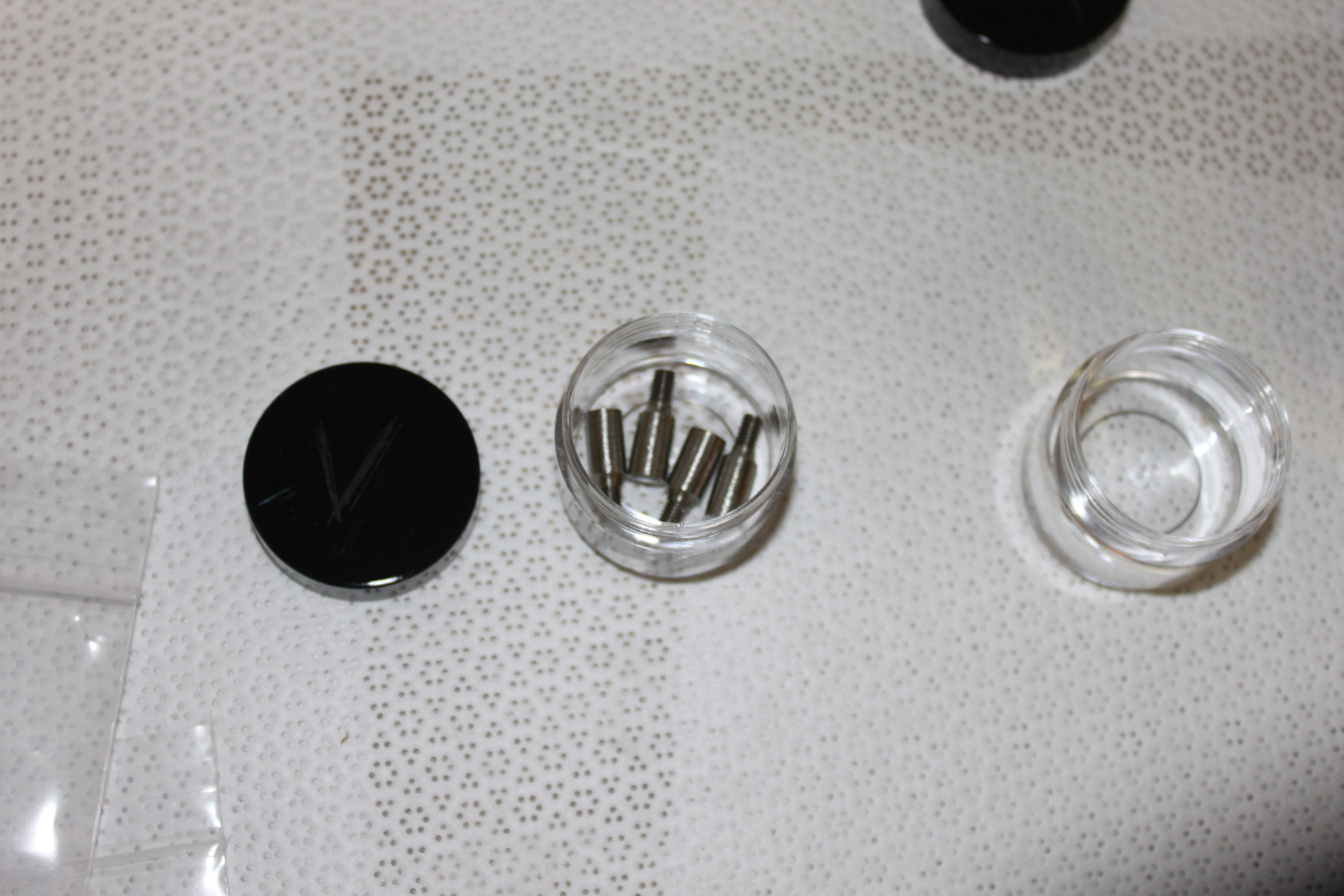}
\quad
\includegraphics[width=.30\textwidth]{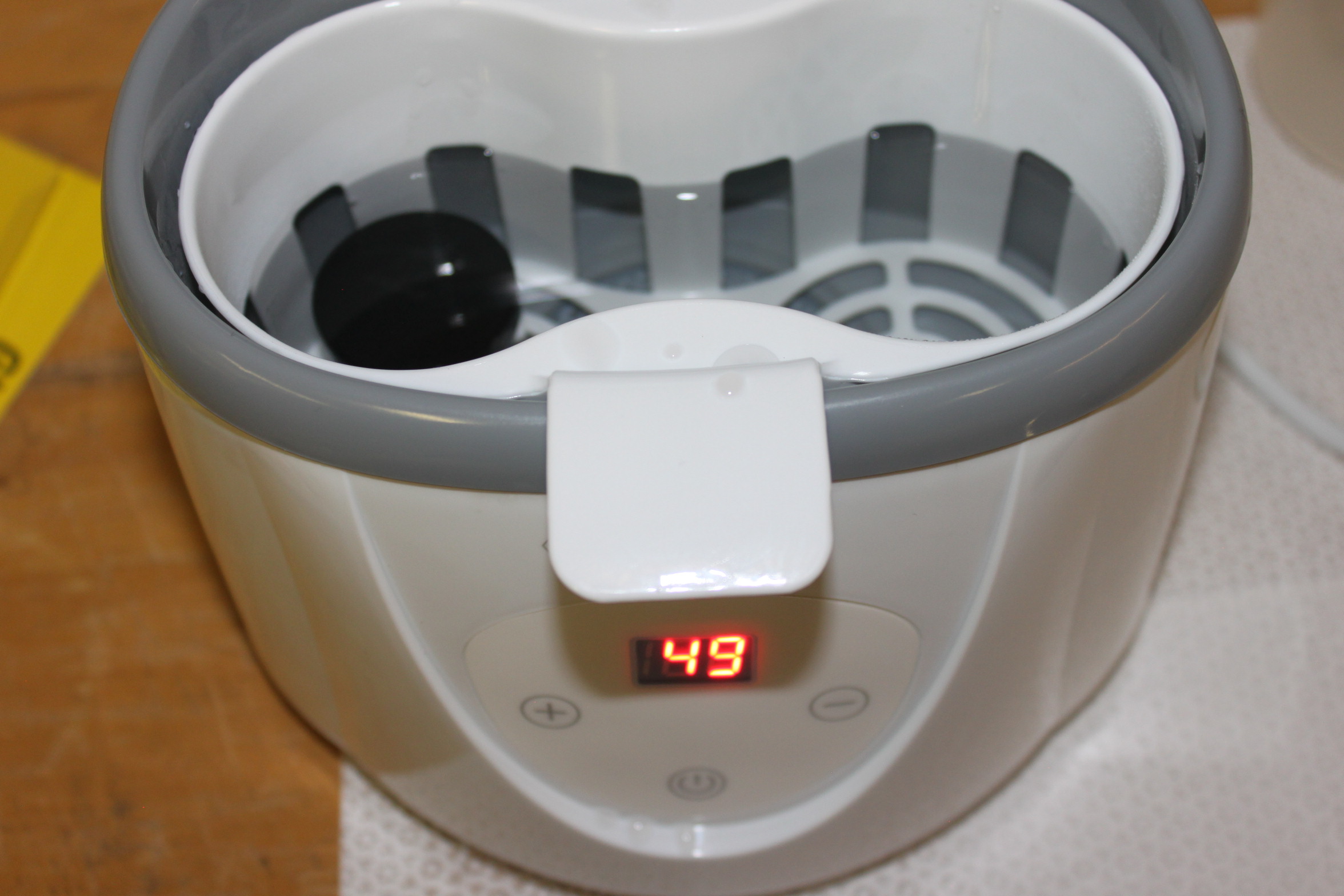}
\quad
\includegraphics[width=.30\textwidth]{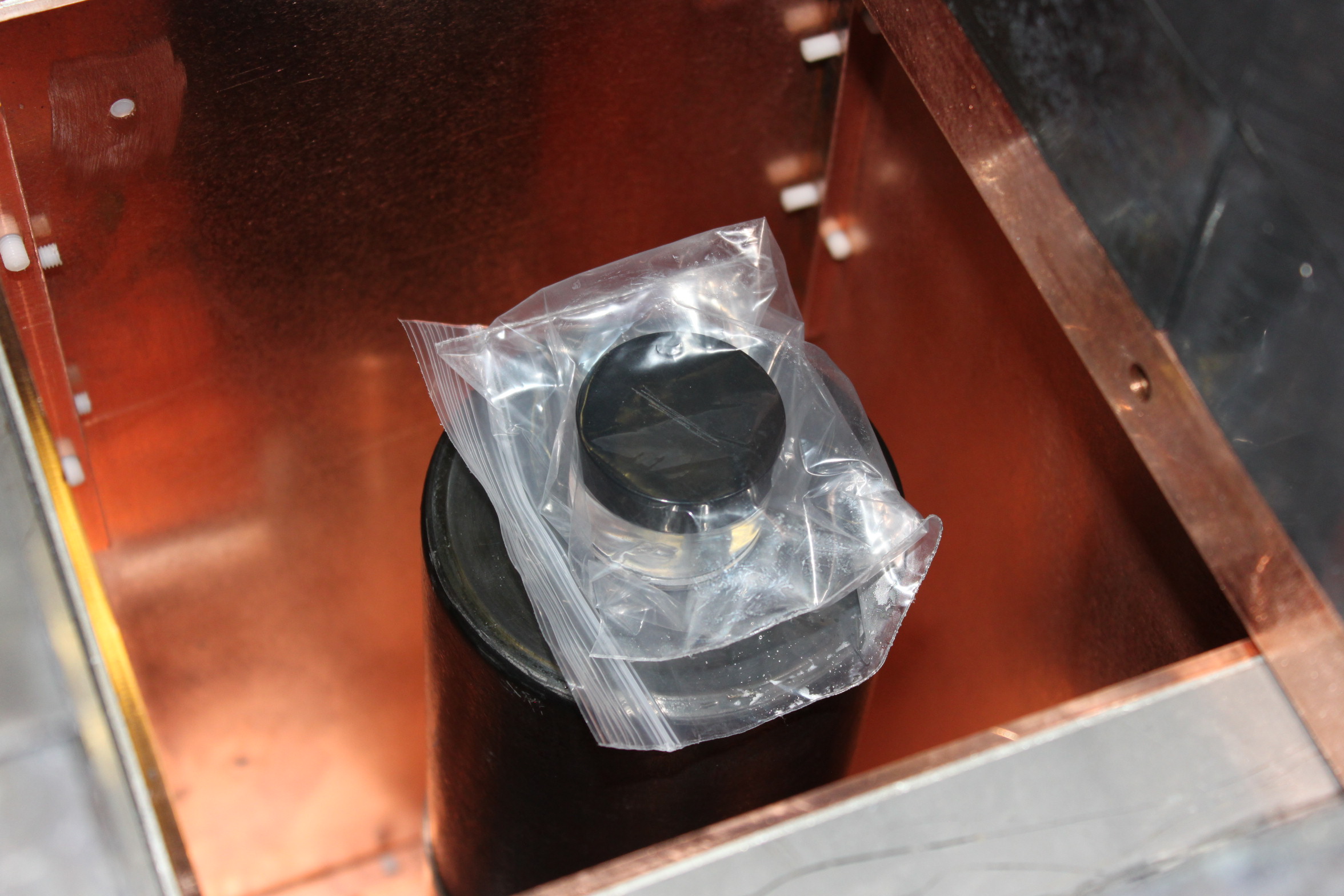}
\caption{\label{Fig_Cryo} (Left) Four source capsules after cryogenic exposition in a water container. (Middle) Water container in a controlled temperature environment. (Right) Water container on top of a germanium low-background counter.}
\end{figure}

After welding, the sources are considered encapsulated and sealed sources. In order to verify the durability of the encapsulation, especially under cryogenic conditions as present in LEGEND, a procedure following Ref.~\cite{Baudis2015} and the Code for Federal Regulations (CFR)~\cite{CFR} was implemented. The sources were exposed to cryogenic temperatures for ten minutes by dipping them in a dewar filled with liquid nitrogen. After a warm-up period, the sources are wiped following standard Radiation Safety protocols, and the wipes were analyzed using portable $\alpha$ and $\gamma$ counters. No radioactivity was detected. Following the guidance from code 49 CFR 173.469\footnote{corresponds to EU regulation ISO 2919:2012 and ISO 9978:1992}, the source capsules were then placed into water containers in sets of four capsules. Each water container contained 50\,mL of water. No bubbling was observed at the container surfaces. Over a 96-hour period the water was kept at 122$^{\circ}$F (50$^{\circ}$C) with the capsules inside the water container, see Fig.~\ref{Fig_Cryo}. After this period, the source capsules were removed. The exposed water samples were then measured on top of a shielded low-background germanium counter. For comparison,  a water sample was prepared following the same procedure, but with an empty source assembly. 

\begin{figure}[htbp]
\centering 
\quad
\includegraphics[width=.8\textwidth]{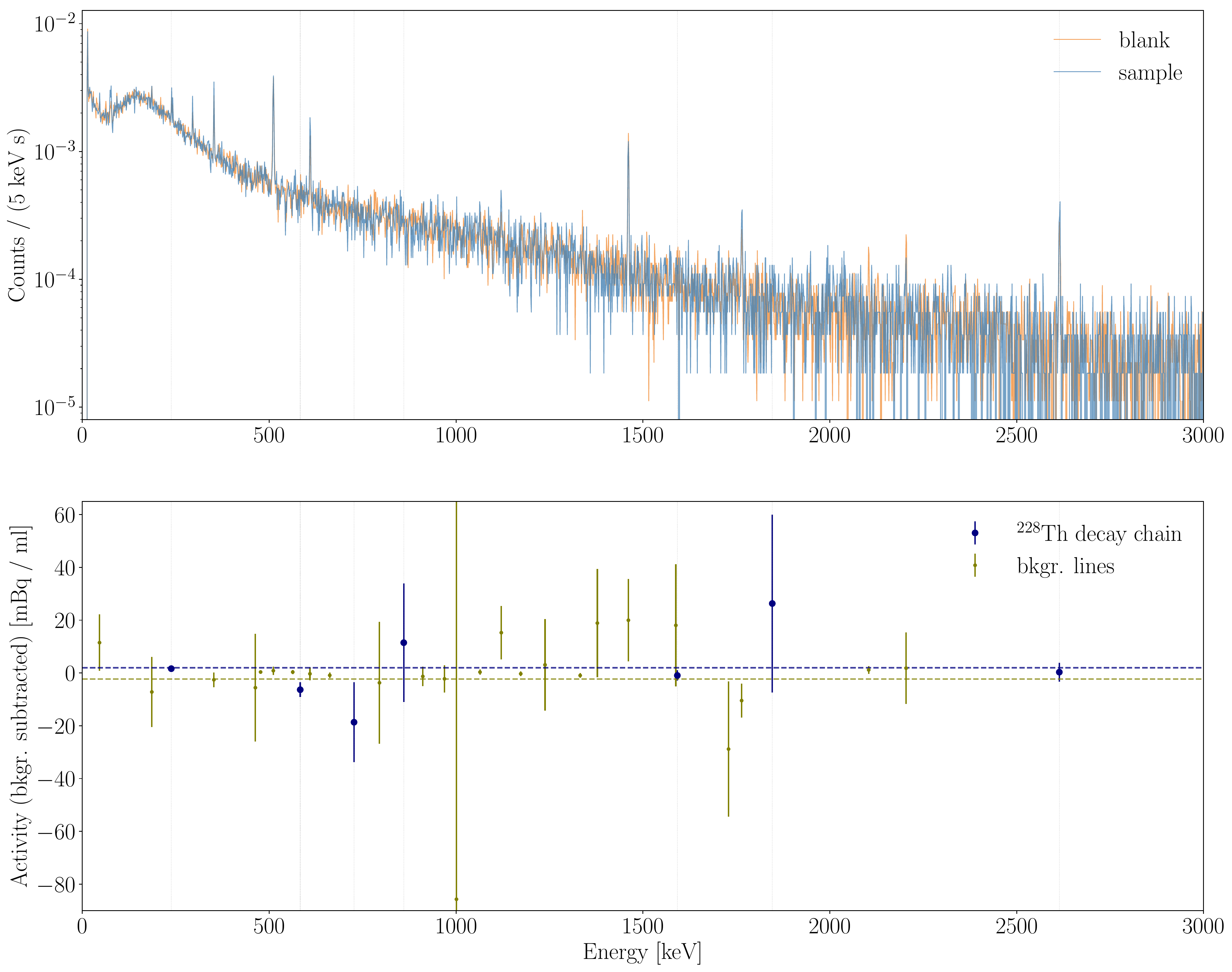}
\caption{\label{Fig_Water} (Top): Spectrum of one water sample that was exposed to the sources, and a sample which used a blank source assembly. (Bottom): Background subtracted content in prominent background signatures (green) and characteristic peak from the $^{228}$Th decay chain (blue). The dashed lines indicate the corresponding sample averages.}
\end{figure}

The low-background counter has an average background rate of about $10^{-4}$\,counts/(keV$\cdot$s) in the region between 100 and 2700 keV. For the 238-keV $\gamma$ ray emitted within the $^{228}$Th-decay chain, the detection efficiency is about 30\%. This results in a detection sensitivity of 3\textendash5 mBq/mL. This sensitivity by far exceeds the requirements of the CFR regulation, which is 2\,kBq. However, given the extremely low background conditions required for LEGEND, the improved sensitivity is helpful to ensure that no leakage endangers the experiment. As Fig.~\ref{Fig_Water} shows, the observed spectrum with a water sample is consistent with the background expectation. No intensity in prominent transitions of the $^{228}$Th decay were observed in any water sample. Hence, it can be concluded that the source container passed the seal test under cryogenic conditions. 

\section{Source characterization}
\label{sec:source_characterization}
$^{228}$Th sources with an activity of approximately 5\textendash6 kBq each are required to provide uniform and sufficient statistics for the germanium detectors in \Ltwo{}. At the same time, contributions to the background radioactivity should be minimized. This section shows the characterization of all sources, including their activities determined by $\alpha$ and $\gamma$ measurements, and the neutron fluxes yielded from $\left(\alpha,n\right)$ reaction.

\subsection{Activity determination}
\label{sec:activity_determination}

\begin{figure}[htbp]
\centering 
\includegraphics[width = 0.75\textwidth]{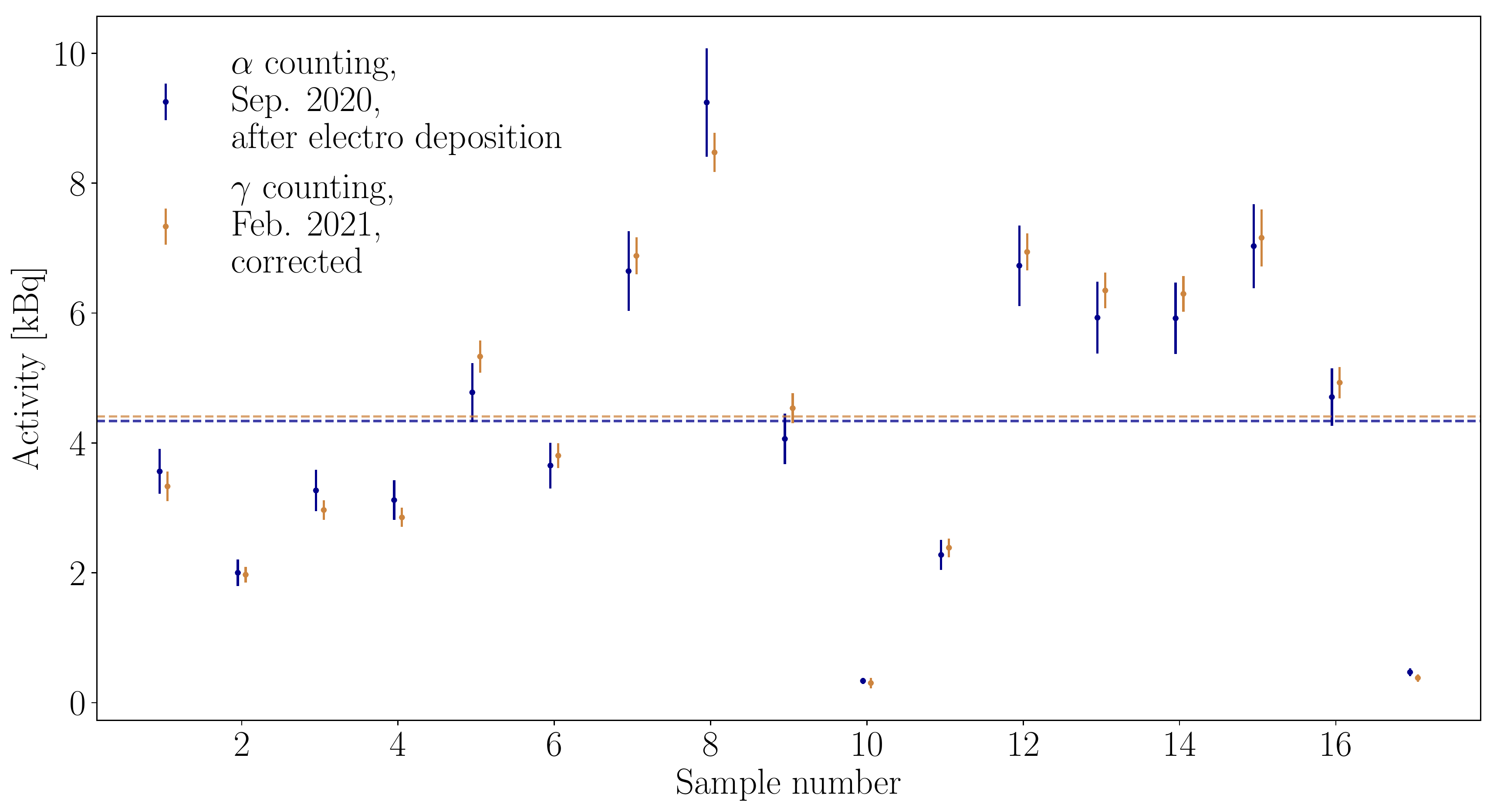}
\caption{Comparison of activity values as measured with different methods. After electrodeposition, the $\alpha$ activity of the samples was determined using a silicon surface barrier detector (blue). Before shipping, the activity after all manufacturing was determined with a high-purity germanium counter (orange). The dashed lines indicate the corresponding sample means. The small offset between the data points within one sample is for visualization purposes only.}
\label{Fig_Activity}
\end{figure}

Shortly after electrodepositing the material onto the gold foil, the activity was determined with a silicon detector by analyzing the $\alpha$-emission intensity from the initial $^{228}$Th decay. This method is a valuable cross check since it allows a comparison after the full production. Large leaks and loss of activity should be visible by comparing the initial $\alpha$ counting with a $\gamma$ counting after production. Therefore, all sources were measured individually with a high-purity germanium detector. Each individual source was placed at a 10-inch distance from a germanium detector. For this detector and geometry, the full-energy-peak efficiency of the 2614.5-keV transition was measured with a known calibrated source and determined to be \num{5.5(2)e-4} within a 20-keV-wide window around the peak center. This efficiency describes the ratio of measured to emitted $\gamma$-rays.  A similar window was used for the spectra of the LEGEND sources. The width of the 20-keV wide window was chosen from the energy resolution of the detector, which was around 4-keV, and the fact that Compton scattered events should be avoided. After subtraction of a background spectrum, measured over 3 hours without a source present, all counts in this window can be assumed to originate from the 2614.5-keV transition. The transition has within the decay chain a 35.6\% emission probability. In thirty-minute-long measurements, the activity can be determined on the 2-10\% level, which is comparable to the standard precision of commercially certified sources. The activity (Fig.~\ref{Fig_Activity}) shows that after correction for the time delay due to the production time, the activities using two different methods are in agreement with each other. This extra delay of about three months also guaranteed that the daughter isotopes within the decay chain are in equilibrium with the $^{228}$Th decay itself. As one can see in Tab.\,\ref{tab:i} some sources ended up in a low deposition and are considered low-activity sources. They are in use currently, but will be replaced with a second set of similar sources during the early phase of \Ltwo{}.

\subsection{Neutron emission}

\begin{figure}[h]
\centering 
\hspace{-0.6cm}
\qquad
\hspace{-0.8cm}
\includegraphics[width=.375\textwidth]{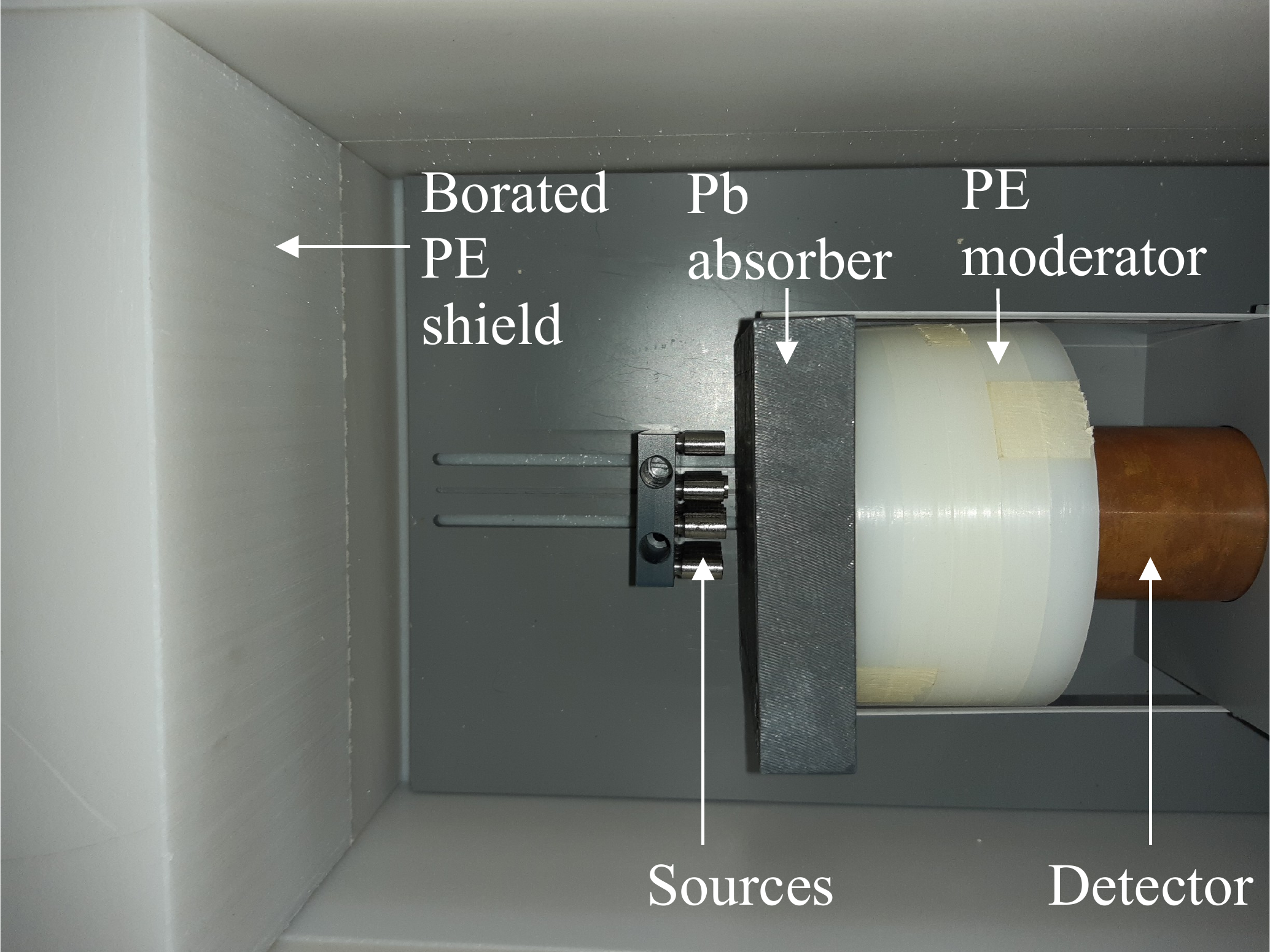}
\qquad
\hspace{-0.8cm}
\includegraphics[width=.205\textwidth]{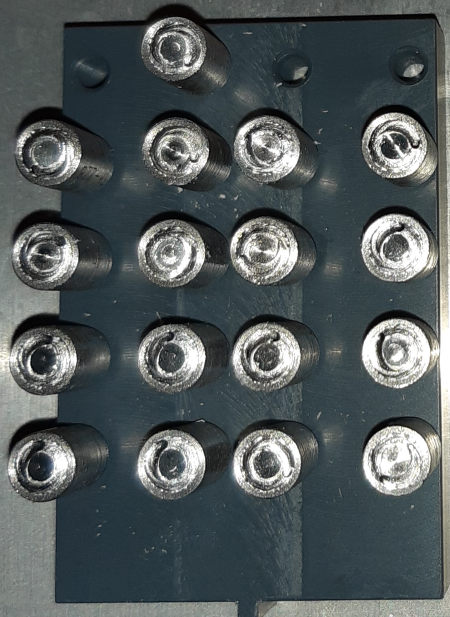}
\caption{(Left): Setup of the LiI(Eu) detector deployed for the measurement of the neutrons emitted by the calibration sources. When neutrons hit the $^{6}$Li target, an $\alpha$ particle and a $^{3}$H ion are produced with a combined energy of $E=4.78$~MeV, exciting surrounding atoms. A PMT detects the secondary scintillation light, which results from the subsequent de-excitation processes. We refer to the text for a detailed description of the setup. (Right): Image of the 17 sources mounted into the source holder for the measurement.} 
\label{fig:LiIdetector}
\end{figure}

Given the potential impact of neutrons on the $0\nu\beta\beta$ sensitivity, we performed a direct measurement of the neutron emission of the calibration sources. We deployed a modified LiI(Eu) detector system enriched to 96\% in $^{6}$Li from SCIONIX\footnote{\url{https://scionix.nl/}}, operated underground at LNGS in a low background environment shielded from external neutrons. The detector consists of a cylindrical crystal placed into a custom-made copper housing. It is equipped with an R8250 photomultiplier tube (PMT) from Hamamatsu Photonics\footnote{\url{https://www.hamamatsu.com/eu/en.html}}, providing an ultra-low background of less than 1 neutron per day, see Ref.~\cite{Benato2015}. The housing is connected to an external HV supply. A detailed description of the detector system can be found in Ref.~\cite{Tarka2012}.\\The radioactive sources were mounted into a PVC holder with 20 source slots in total. A 2-cm-thick Pb block shielded the emitted $\gamma$s, and a 5-cm-thick Polyethylene (PE) layer moderated the emitted neutrons for thermalization. The moderation increases the detection efficiency as the neutron capture cross section scales with $1/v$, where $v$ denotes the velocity of the neutrons. The entire setup was placed in a shield made of borated PE, with on average 5\% boron, in order to moderate and capture environmental neutrons before penetrating the detector. Pictures of the setup and the sources mounted into the PVC holder are shown in Fig.~\ref{fig:LiIdetector}.

\label{sec:neutron_emission}

Incoming neutrons interacting inside the $^{6}$Li crystal cause the production of an $\alpha$ particle and a $^{3}$He ion, with a combined energy of $E=4.78$~MeV. This emission of the ions leads to the excitation of surrounding atoms. The PMT detects the secondary scintillation light from the subsequent de-excitation processes. The PMT had been operated at 750~V, a voltage providing both sufficient gain and an acceptable width of the neutron peak of $~$9.5\%, as determined with the help of spectra acquired with an $^{241}$Am$^{9}$Be source of activity 160(4)~n/s in 2013, see ~\cite{PhysRevD.88.012006}. The data acquisition had been performed with a stand-alone ORTEC multi-channel analyzer\footnote{\url{https://www.ortec-online.com/products/electronics/multichannel-analyzers-mca}}. Recorded voltage signals were converted into event channel energies via a trapezoid filter with a flat-top and a risetime of \SI{2}{\micro s} each.


\begin{figure}[h]
\centering 
\includegraphics[width = 0.75\textwidth]{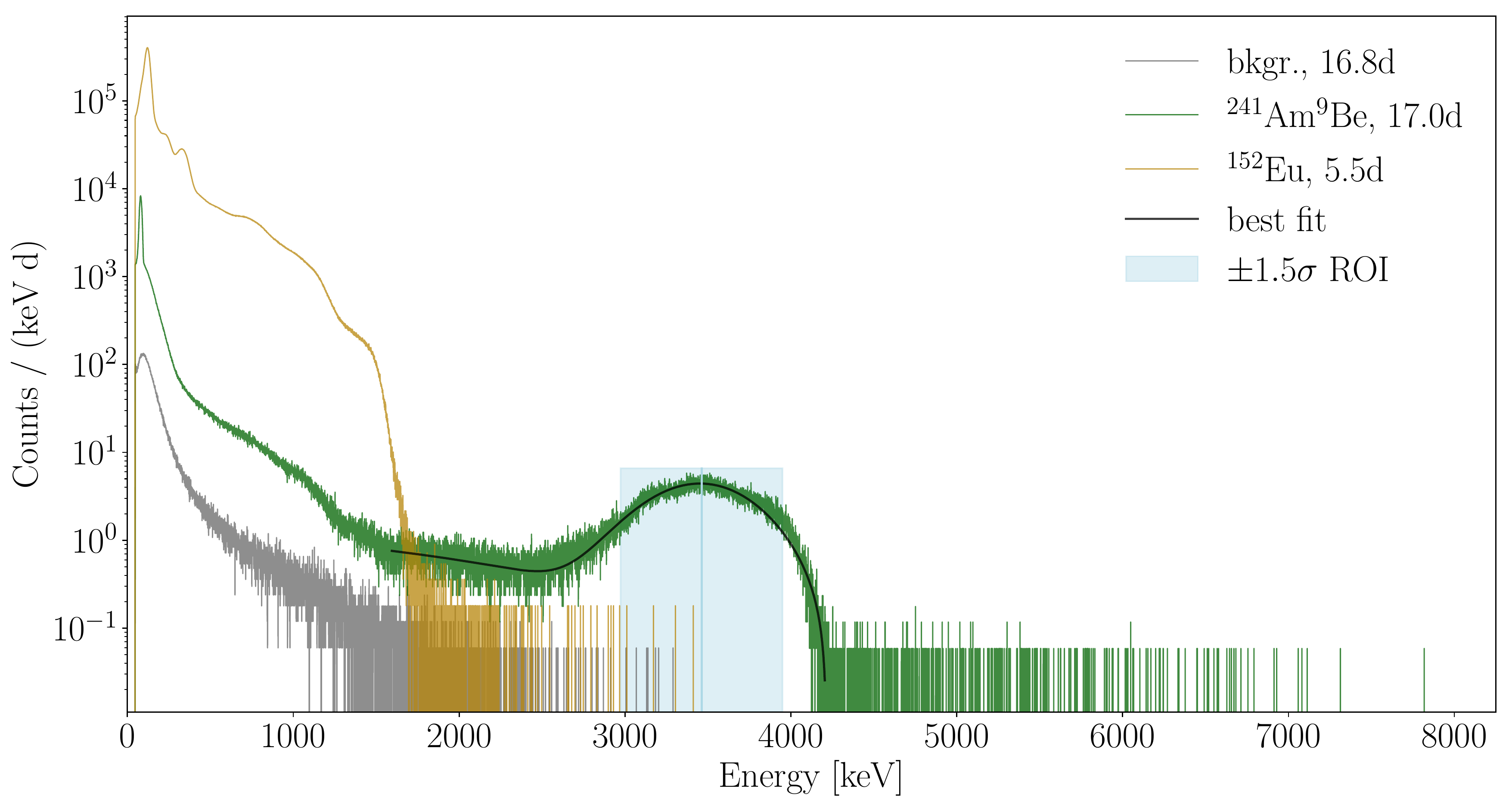}
\caption{
Spectra as measured with an $^{152}$Eu $\gamma$ source for energy calibration (gold), with an $^{241}$Am$^{9}$Be n source for efficiency determination (green), and without a source (gray). The marked region corresponds to a $\pm 1.5 \sigma$ region of interest for thermal neutrons as determined as the standard deviation of the Gaussian component of the fit of the n peak (shown in black) as outlined in the text.} 
\label{fig:n_eff_calib}
\end{figure}

\begin{figure}[h]
\centering 
\includegraphics[width=.75\textwidth]{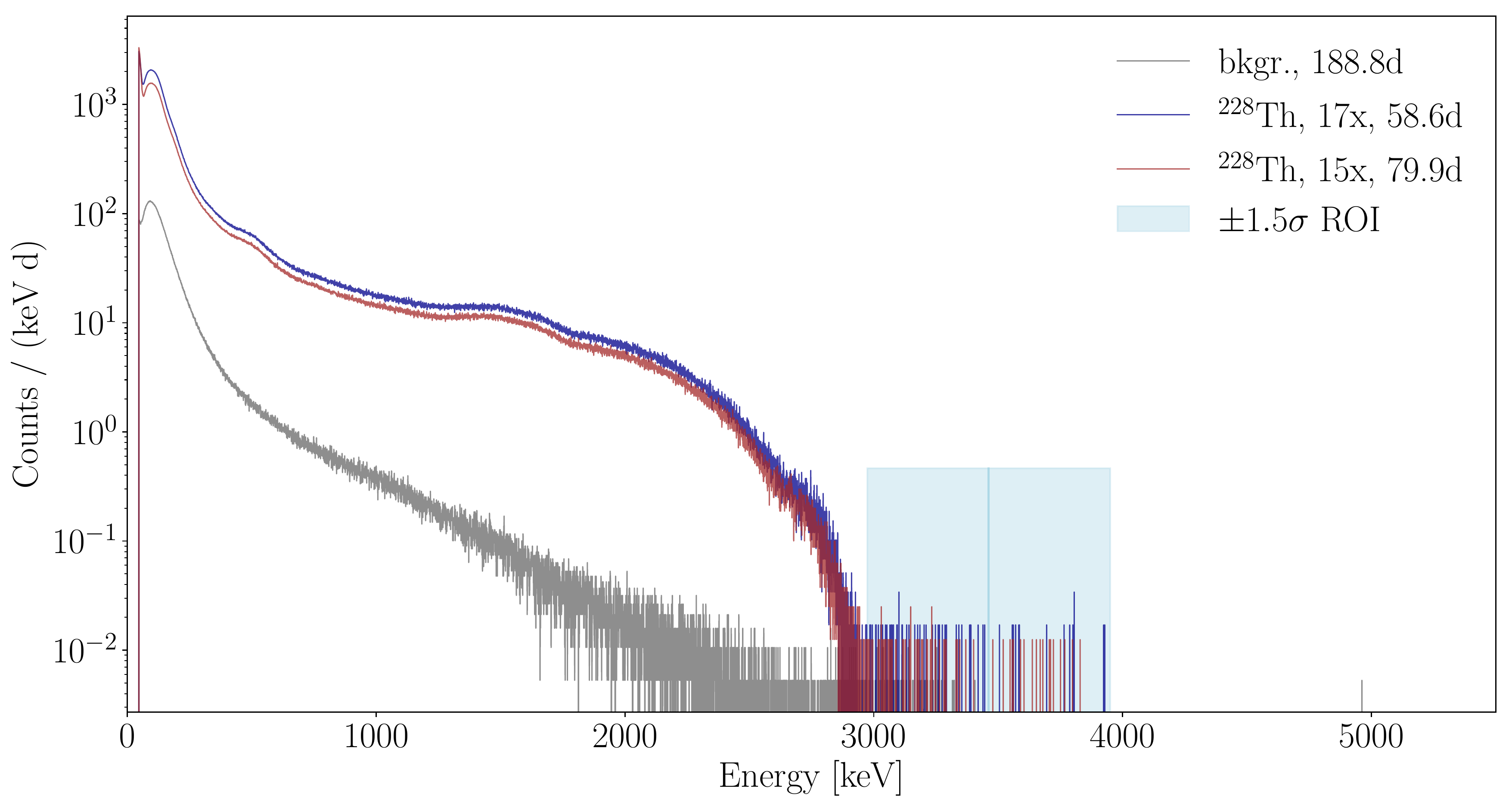}
\caption{Neutron spectra as obtained from two measurements with 17 (blue) and 15 (red) calibration sources, respectively. The background spectrum (gray) is shown as well. The marked light blue region was determined in Fig.~\ref{fig:n_eff_calib}, which indicates the region of interest for thermal neutrons used for the neutron-flux estimation.} 
\label{fig:n_measurement}
\end{figure}

To prepare the neutron measurement, we calibrated the energy scale of the detector by accumulating 5 days of data with an $^{152}$Eu $\gamma$ source. The corresponding linear calibration curve was deduced from the position of known $\gamma$-ray lines in the $^{152}$Eu spectrum, which is shown in Fig.~\ref{fig:n_eff_calib}.
Furthermore, we determined the neutron capture detection efficiency of the LiI(Eu) detector by deploying the $^{241}$Am$^{9}$Be source introduced above in a 17-day measurement shown in Fig.~\ref{fig:n_eff_calib} as well, which we compared to equally long background only data taking. Note that, in contrast to the measurements described in~\cite{Benato2015}, these data had been taken underground. This strongly reduced the environmentally induced neutron background to a negligible number of 12 events. The neutron detection efficiency was deduced from a Gaussian fit of the neutron peak on top of the continuum. The latter was modelled as a linear and two step functions, a logistic curve plus a complementary error function. Two step functions were needed to obtain a smooth transition of the continuum from above to below the signal peak. The $\chi^2$ over degrees-of-freedom (dof) statistic yields a goodness-of-fit estimate of 1.4. The efficiency as estimated from the amplitude of the Gaussian peak is determined to be $\epsilon = 4.55(3) \times 10^{-4}$, approximately 14\% lower than the value measured above ground, cf.~\cite{Benato2015}. Here the digit in brackets denotes the statistical error of the fit. Systematic uncertainties are given by the fit range inducing approximately 6\% of uncertainty as estimated via different fit ranges, the flux uncertainty contributing another 2.5\%, and the uncertainty due to the fit method itself. To estimate the fit method-induced contribution, we also performed a counting experiment for comparison. We assumed Poisson distributions for the number of signal and background events, and constrained the neutron flux by a Gaussian distribution with a standard deviation of 4~n/s. The maximum likelihood estimate for the efficiency in this alternative determination is given as $\epsilon_{\rm alt} = 4.50(1) \times 10^{-4}$ to $5.02(2) \times 10^{-4}$, depending on the choice of the counting range within $\pm$2 to 4$\sigma$ intervals. As mentioned above, we observe 12 events only in the former window, which is fully irrelevant for this estimation considering a best-fit amplitude of $\mathcal{O}(10^5)$ events. Due to the slightly deviating values we add an additional systematic contribution of 9\% to compensate for the effect of different counting ranges and analysis methods.\\
A further, geometry related impact on the efficiency estimation needs to be taken into account. In the calibration source measurements all available sources were measured together to reduce the data-acquisition time required to collect sufficient statistics. In these measurements, the sources were mounted in different source slots on the holder, whereas for the efficiency determination only one central slot was required. This discrepancy in the settings induces further geometrical complications, more precisely, different moderation lengths and solid angles. The former was estimated with a second efficiency measurement deploying only a PE moderator of \SI{3}{cm} thickness (instead of \SI{5}{cm}). A fit of the new peak amplitude yielded a decrease of 11\%, which was taken into account as a conservative estimation of this systematic uncertainty on the efficiency. To compensate for the latter effect, we applied the solid angle correction for a cylindrical detector geometry with point-like sources outlined in \cite{PRATA2003599}, obtaining a mean effective solid angle of 0.97, with a standard deviation of 0.05. Due to both the difficulties of measuring each geometrical source slot-to-crystal distance inside the setup accurately, and also the fact that the radioactive material is distributed within a volume and not concentrated at a single point, we further appended an additional systematic uncertainty of 3\% to the solid angle. This value was estimated by varying the distance by 4~mm, the length spanning the space in which the radioactive material is distributed, see Fig.~\ref{fig:SIS}, right. To analyse the final neutron measurement, we estimated the effective efficiency as a source-activity- and solid-angle-weighted average over all source contributions. That is, we defined the weighted efficiency as
\begin{equation}
\epsilon_{\rm w} = \epsilon \sum_{i=1}^{17 \left[ 15 \right]} \frac{A_{i}}{A_{\rm tot}} \frac{\Omega_i}{\Omega_\epsilon} ~,
\end{equation}
where $\Omega_i$ denotes the solid angles of the source threads. We normalized the solid angles with respect to the thread used for the efficiency measurement $\Omega_\epsilon$. The individual activities $A_i$ were normalized to the total activity $A_{\rm tot}$ of all sources combined. We determined the total systematic uncertainty of the effective weighted efficiency as the square root of the quadratic sum of all contributions mentioned above, including the weighing of the solid angle and the activity. To estimate the neutron flux, we took data for 58.6 days with 17 sources, and continued another 79.9 days with 15 sources. Two sources were removed earlier to perform LEGEND-200 related calibration measurements. The background was measured for 188.8 days. The measured spectra are shown in Fig.~\ref{fig:n_measurement}. 

Due to the Compton continuum at energies below the thermal-neutron-signal region, we constrained the search range to a $\pm 1.5 \sigma$ interval, where the resolution was taken from the efficiency measurement. We obtained the following event numbers in the region-of-interest (ROI): $n=$~60 events with 17 sources, $n=$~77 events with 15 sources, and $b=$~73 events in the background data. We applied again a counting statistic under the assumption of Poisson-distributed-event counts, combined with a Gaussian distribution to constrain the solid angle and activity weighted efficiency $\epsilon_{\rm w}$. A second Gaussian term models the combined, total source activity $A'$, after correcting for the decrease due to the exponential decay since the characterization of the source activities as described in Sec.~\ref{sec:activity_determination}. Minimising the likelihood function for the counting statistic yields the following neutron-flux estimates for the two measurements,
\begin{equation}
\Psi = \frac{n - t_{\rm s} / t_{\rm b} b}{\epsilon_{\rm w} A' t_{\rm s} f_{\rm ROI}} = \begin{cases} 4.00 \pm 0.76_{\rm stat} ~ \times 10^{-4} ~\text{n / (kBq$\cdot$s)} & ~17~\text{sources}\\ 4.96 \pm 0.89_{\rm stat} ~ \times 10^{-4} ~\text{n / (kBq$\cdot$s)} & ~15~\text{sources} \end{cases}\quad.
\end{equation}
Here, the observed event numbers for the respective measurement $n$, and the background $b$, are corrected for the 1.5$\sigma$ ROI search region by the factor $f_{\rm ROI} \sim 0.8664$. The times $t_{\rm s/b}$ denote the duration of the source and the background measurement, respectively. The statistical uncertainties are determined as a $1\sigma$ confidence interval centered at the maximum likelihood estimator. We estimated the individual systematic uncertainties by propagating the errors of source activity, weighted efficiency, and an additional term compensating for the difference in the neutron source emission spectra of $^{228}$Th and $^{241}$Am$^{9}$Be. This uncertainty had been estimated in~\cite{Maneschig2012} via simulations, yielding 12.1\%. Overall we obtained systematic uncertainties of $0.89~ (1.23)~ \times 10^{-4}$ n / (kBq$\cdot$s) for the 17 (15)-source run, dominating the uncertainty. To determine the global neutron-flux result, we calculated an inverse uncertainty-weighted average of the two individual estimates, including correlations in the data as described in Chapter 4 of Ref.~\cite{Erler:2015nsa}. Given the strong overlap between the measured source configurations, we approximate the correlated, globally present contribution to the systematics directly as $c = s_{\min}$, where $s_{\min}$ denotes the minimum individual systematic uncertainty. Following the prescription in \cite{Erler:2015nsa}, we obtain an estimated correlation coefficient for the systematic uncertainty of $\rho = c^2 /(t_1 t_2) = 0.45$. Here the $t_i^2$ are the sum of squares of statistical and systematic errors. The final, combined neutron flux, under the assumption of approximately Gaussian uncertainties, is evaluated to be 
\begin{equation}
    \Psi = \left( 4.27 \pm 0.60_{\rm stat} \pm 0.92_{\rm syst} \right) \times 10^{-4} ~\text{n / (kBq$\cdot$s)}.
\end{equation}
For comparison, when neglecting the correlations in the data, the direct inverse uncertainty-weighted average yields $\Psi = \left( 4.36 \pm 0.58_{\rm stat} \pm 0.73_{\rm syst} \right) \times 10^{-4} ~\text{n / (kBq$\cdot$s)}$, which slightly underestimates the uncertainties. We have seen that the dominating uncertainty is of systematic nature. In particular, the efficiency determination with the single $^{241}$Am$^{9}$Be neutron source yields a slightly different neutron emission spectrum compared to the multiple $^{228}$Th sources. Also, the source slots at the holder lead to different solid angles as well as moderation lengths. A more precise evaluation may thus be obtained by either determining the efficiency for each of the threads in the source holder separately, or by measuring each of the sources individually. However, such measurements would require several year-long measurement campaigns. We would like to state that the neutron flux determined is of the same order as the flux emitted by the GERDA II calibration sources, see \cite{Benato2015, Miloradovic2020}, but slightly lower, even when considering the uncertainties. Compared to commercial sources, the neutron flux is reduced by approximately an order of magnitude, see \cite{Tarka2012,Benato2015}. We conclude that the new sources provide sufficiently low neutron-emission properties, proving the possibility to safely deploy these calibration sources in the LEGEND-200 experiment. A precise estimate of the background contribution induced in LEGEND-200 by calibrations had been determined via simulations within the LEGEND program directly. Assuming a reasonable schedule of weekly calibrations with an average duration of 2 hours, an acceptably low calibration-induced background level of around $\mathcal{O}\left(10^{-5}\right)\left(\mathcal{O}\left(10^{-7}\right)\right)$~cts~/~(keV$\cdot$kg$\cdot$yr) before (after) background reduction cuts had been found. 

\section{Summary and outlook}
\label{sec:summary_outlook}

The \LEG{} experiment employs enriched high-purity germanium detectors to search for \onbb{} with a half-life sensitivity beyond 10$^{28}$\,years.
For the operation of the germanium detectors, it is crucial to determine and monitor the energy scale and the energy resolution of the detectors. In the \Ltwo{} experiment, four calibration systems, equipped with four $^{228}$Th sources each, are deployed regularly to calibrate the detectors' energy responses and to measure their resolutions. Sources with a low neutron emission rate are required to avoid activation of the germanium crystals. In this paper, the production of the sources and the characterization measurements of the source properties are described in detail.

$^{228}$Th in the form of thorium-nitrate dissolved in HCl was electrodeposited on a 50-\si{\micro m}-thin gold foil. The high $\left( \alpha, n \right)$ energy threshold of gold prevents the production of neutrons, which reduces the neutron flux in comparison with commercial Th sources embedded in ceramic materials. The length of the deposition was adjusted based on an $\alpha$-activity measurement of a test sample immediately after the deposition, to reach the desired activity of $\sim$5~kBq per source. After deposition, the foils were rolled and placed inside doubly encapsulated source containers sealed with electron beam welding in vacuum. The durability of encapsulation was tested in two steps, both giving no indication of radioactivity leakage.

To accurately quantify the source activities, two measurement approaches were performed. First, a silicon detector was used to measure the $\alpha$-emission intensity of the $^{228}$Th decays right after the electrodeposition. Afterwards, all sources were measured individually with a high-purity germanium detector, whose efficiency at the full-energy peak of the \SI{2614.5}{keV} transition was determined by a known calibrated source. The activities could be constrained to an accuracy level of 2-10\% in thirty-minute measurements. After correction for the time delay due to the production time, both measurements gave consistent activity estimations, and are on the same order of magnitude as the expected source activities.
A direct measurement of the neutron emission rate of the sources was performed with a modified, low-background LiI(Eu) detector operated in a neutron shield underground at LNGS.\@ The energy scale of the detector was calibrated with a $^{152}$Eu $\gamma$ source. Its neutron capture detection efficiency was determined with an $^{241}$Am$^{9}$Be source, yielding $4.55 \times 10^{-4}$. The neutron fluxes were measured in two measurements with all sources available, leading to a combined result of $\Psi = \left( 4.27 \pm 0.60_{\rm stat} \pm 0.92_{\rm syst} \right) \times 10^{-4} ~\text{n/(kBq$\cdot$s)}$. The neutron background in \Ltwo{} as induced by calibration runs was estimated to be $\mathcal{O}\left(10^{-5}\right)\left(\mathcal{O}\left(10^{-7}\right)\right)$~events~/~(keV$\cdot$kg$\cdot$yr) before (after) background reduction cuts. Compared to the LEGEND-200 background goal of $\mathcal{O}(10^{-5})$~events~/~(keV$\cdot$kg$\cdot$yr) after cuts, \cite{LEGEND:2021bnm}, the estimated value is considered sufficiently low to safely deploy the radioactive sources in the experiment. 
Radioactive sources are required to calibrate particle detectors, yet minimizing the background contribution originating from the sources themselves is indispensable for rare event searches. We demonstrated a procedure to produce low-neutron-background $^{228}$Th sources to be deployed in the \Ltwo{} experiment, which can be extended to other future experiments requiring low-neutron-emitting radioactive sources.

\section{Acknowledgments}
The authors thank A.~Duffield and S.~Wiest and their team at LANL Sigma division for making the welding possible at LANL. We also thank C.~Ransom, W.~Xu and L.~Paudel for their help with Monte Carlo simulations, as well as M.~Laubenstein for his assistance with the source approval at LNGS and with the neutron flux measurement. We thank J.~Franchi for drawings and hardware support, and the UZH workshop for the source holders and mechanics production. We also thank the \LEG{} hardware team for helping to install the calibration systems and the sources in the experiment. This work was supported by the Department of Energy, Office of Nuclear Physics under Federal Prime Agreement LANLEM78, by the Swiss National Science Foundation under Grants No. 200020-204950 and 20FL20-201537 and by the Candoc Grant No. K-72312-09-01 from the University of Zurich. 

\appendix

\section{Activity table}

\begin{table}[htbp]
\centering
\caption{\label{tab:i} Activity of the individual source assemblies as of Feb. 9, 2021}
\smallskip
\begin{tabular}{cc}
\hline
Source \#No& Activity [Bq]\\
\hline
1 & 2922 (133 ) \\
2 & 1728 (72 ) \\
3 & 2603 (88 ) \\
4 & 2502 (86 ) \\
5 & 4675 (146 ) \\
6 & 3337 (113 ) \\
7 & 6034 (166 ) \\
8 & 7431 (175 ) \\
9 & 3977 (136 ) \\
10 & 266 (46 ) \\
11 & 2093 (84 ) \\
12 & 6087 (167 ) \\
13 & 5567 (162 ) \\
14 & 5521 (160 ) \\
15 & 6278 (257 ) \\
16 & 4322 (141 ) \\
17 & 335 (33 ) \\
\hline
\end{tabular}
\end{table}

\newpage 

\bibliographystyle{JHEP}
\bibliography{LEGEND200}

\end{document}